\documentstyle[12pt]{article}

\newcommand{\EQ}{\begin{equation}}
\newcommand{\EN}{\end{equation}}

\begin{document}

\topmargin 0pt
\oddsidemargin 5mm
\newcommand{\NP}[1]{Nucl.\ Phys.\ {\bf #1}}
\newcommand{\PL}[1]{Phys.\ Lett.\ {\bf #1}}
\newcommand{\NC}[1]{Nuovo Cimento {\bf #1}}
\newcommand{\CMP}[1]{Comm.\ Math.\ Phys.\ {\bf #1}}
\newcommand{\PR}[1]{Phys.\ Rev.\ {\bf #1}}
\newcommand{\PRL}[1]{Phys.\ Rev.\ Lett.\ {\bf #1}}
\newcommand{\MPL}[1]{Mod.\ Phys.\ Lett.\ {\bf #1}}
\newcommand{\JETP}[1]{Sov.\ Phys.\ JETP {\bf #1}}
\newcommand{\TMP}[1]{Teor.\ Mat.\ Fiz.\ {\bf #1}}

\renewcommand{\thefootnote}{\fnsymbol{footnote}}

\newpage
\setcounter{page}{0}
\begin{titlepage}
\begin{flushright}
SISSA-92-EP-4
\end{flushright}
\vspace{0.5cm}
\begin{center}
{\large The two-dimensional O(3) nonlinear $\sigma$-model at finite temperature
} \\
\vspace{1cm}
\vspace{1cm}
{\large  M\'arcio Jos\'e  Martins
\footnote{on leave from Departamento de Fisica, Universidade Federal de
S.Carlos, C.P. 676 - S.Carlos 13560, Brazil.}
\footnote{martins@itssissa.bitnet}} \\
\vspace{1cm}
{\em International School for Advanced Studies \\ Strada Costiera 11\\
34014, Trieste, Italy } \\
\end{center}
\vspace{1.2cm}

\begin{abstract}
We present a direct derivation of the thermodynamic integral equations of
the O(3) nonlinear $\sigma$-model in two dimensions.
\end{abstract}
\vspace{.5cm}
\centerline{January 1992,~~Phys.Lett.B,~in press}
\vspace{.3cm}
\end{titlepage}

\renewcommand{\thefootnote}{\arabic{footnote}}
\setcounter{footnote}{0}

\newpage
The two-dimensional O(3) nonlinear $\sigma$-model is a well known example
of an exactly integrable field theory  \cite{lu,po2} which is asymptotically
free at short distances and a mass gap appears in the spectrum
through the dimensional
transmutation mechanism \cite{po1}. Its action $A$ is defined by
\EQ
A= \frac{1}{g} \int d^2 x {({\partial}_{\nu} \vec{\eta})}^2
\EN
where $\vec{\eta}=(n_x,n_y,n_z)$ is a three component unit vector on the sphere
$S^2$; ${\vec{\eta}}^2=1$.

It is believed that the O(3) nonlinear $\sigma$-model has no bound states and
its fundamental particles belong to the O(3) multiplet. Using these hypotheses
and the factorized bootstrap A. Zamolodchikov and Al. Zamolodchikov \cite{zz}
have found the respective two-body S-matrices $S_{ij}^{kl}(\beta)$;
\EQ
S_{ij}^{kl}(\beta)=\frac{\pi+ i \beta}{(2 \pi +i \beta)(\pi -i \beta)}
\left ( \frac{2 \pi i \beta}{\pi +i \beta} \delta_{i,j} \delta_{k,l}
-2 \pi \delta_{k,j} \delta_{i,l} -i \beta \delta_{k,i} \delta_{l,j} \right )
\EN

The solution for
the O(3) nonlinear $\sigma$-model directly
from its bosonic representation (1) is still unknown. However,
Wiegmann \cite{wi}, using a
fermionic formulation proposed early by Polyakov and Wiegmann \cite{powi}, was
able to find the associated Bethe ansatz equations from a subtle limit of
an exactly solved four fermion model. These equations are given by,
\EQ
e^{iLm sh(\beta_i)}= \prod_{j=1,i \neq j}^{N} \frac{\beta_i-\beta_j+i \pi}
{\beta_i-\beta_j -i \pi} \prod_{j=1}^{M} \frac{\beta_i-\lambda_j+i \pi}
{\beta_i-\lambda_j-i \pi},~~ i=1,2,...,N
\EN
\EQ
\prod_{j=1,i \neq j}^{M} \frac{\lambda_i-\lambda_j+i \pi}
{\lambda_i-\lambda_j -i \pi} \prod_{j=1}^{N} \frac{\beta_j-\lambda_i+i \pi}
{\beta_j-\lambda_i-i \pi},~~ i=1,2,...,M
\EN
where L is the size of the system, m is the mass of the particles $A_i,
i=1,2,3$
; and the corresponding energy is $E=\sum_{i=1}^{N} m ch(\beta_i)$.

Recently there has been a renewed interest in studying several properties of
model (1). M. L\"uscher and U. Wolff \cite{wolu} have
measured the S-matrices (2)
, exploring the two-particle phase-shift. P. Hasenfratz $et~all$ and
P. Hasenfratz and F. Wiedermayer \cite{ha} have calculated the exact value
of the mass gap m in terms of the renormalization group scale. V. Fateev and
Al. Zamolodchikov \cite{faza} have conjectured the thermodynamic Bethe ansatz
{}~(TBA) equations of the nonlinear O(3) $\sigma$-model as the $M \rightarrow
\infty $ limit from those of the Z(M) parafermionic field theory perturbed
by a certain relevant field that breaks the Z(M) symmetry.

In this letter
we present a direct derivation of the TBA equations of the O(3) nonlinear
$\sigma$-model from Eq.(2). This is a reliable approach and has the advantage
of allowing one to generalize our results in the general case of O(N) symmetry.
We also discuss the presence of a magnetic field coupled
with its conjugate conserved
Noether charge.

In order to deduce such TBA equations, one has first to notice that the
eigenvalue problem for the S-matrices (2) defined by finding a matrix
R($\lambda$)
satisfying the relation,
\EQ
R(\lambda-\nu) T(\lambda) \bigotimes T(\nu)=T(\nu) \bigotimes T(\lambda)
R(\lambda-\nu)
\EN
where the monodromy matrix $T(\lambda)$ is
\EQ
T(\lambda)= S(\beta_1 -\lambda) S(\beta_2-\lambda)....S(\beta_n-\lambda)
\EN
is related with the similar problem solved in the case of the Heisenberg chains
\cite{taba}. Indeed the rapidities $\lambda_j$, satisfying Eq.(4), can be
interpreted as the massless excitations of the spin-1
Heisenberg chain \cite{taba}. This observation allows us to conclude that in
the
thermodymanic limit, $L \rightarrow \infty$, the rapidities $\lambda_j$ form
strings of the following type \cite{str}:
\EQ
\lambda_j= \lambda_j^m +\frac{i \pi}{2}(m+1-2 \alpha);~~ \alpha=1,2,...,m
\EN
where the $\lambda_j^m$ are real; the center of the m-string $\lambda_j$.

Substituting Eq.(7) in Eqs.(3,4), taking the logarithm and the thermodynamic
limit , we rewrite Eqs.(3,4) as,
\EQ
msh(\beta)=2 \pi Q(\beta) -\int_{-\infty}^{\infty}
\Psi_1(\beta-\beta^{\prime})
\sigma(\beta^{\prime}) d \beta^{\prime}
-\sum_{m=1}^{\infty} \int_{-\infty}^{\infty}
\Psi_m(\beta-\lambda) \rho(\lambda)d \lambda
\EN
\EQ
2 \pi J_m(\lambda) +\sum_{k=1}^{\infty}
\int_{-\infty}^{\infty} \Theta_{mk}(\lambda-\lambda^{\prime})
\rho_k(\lambda^{\prime}) d \lambda^{\prime}= \int_{-\infty}^{\infty}
\Psi_m(\beta-\lambda) \sigma(\beta)d \beta
\EN
where  $\frac{d Q(\beta)}{d \beta}= \sigma(\beta)+\tilde{\sigma}(\beta)$,
$\frac{d J_m(\lambda)}{d \lambda}= \rho_m(\lambda)+\tilde{\rho}(\lambda)$;
$\sigma(\beta)~(\tilde{\sigma}(\beta))$ and
$\rho(\lambda)~(\tilde{\rho}(\lambda))$ are the densities of particles~(holes)
related with the rapidities $\beta_j$ and $\lambda_j$, respectively.
The functions
$\Psi_m(x)$ and $\Theta_{mk}(x)$ are defined by,
\EQ
\Psi_m(x) = g_{m+1}(x) +g_{m-1}(x)
\EN
\EQ
\Theta_{mk}(x)= \sum_{i}( g_i(x)+g_{i+2}(x));~~ i=|m-k|,|m-k|+2,...,m+k-2
\EN
where $g_a(x)=2 artg(\frac{2 x}{\pi a})$ and $g_0(x) \equiv 0$.

The temperature is encoded by using the method pioneered by Yang and Yang
in the case of the delta function interaction \cite{yy}. This technique
consists in the minimization of the free energy $F=E-TS$ in the
equilibrium state. In our model, the energy E and the entropy S are given
in terms of the densities $\sigma(\beta),\tilde{\sigma}(\beta),\rho(\lambda),
\tilde{\rho}(\lambda)$ by,
\EQ
E=\int_{-\infty}^{\infty} mch(x) \rho_0(x) dx
\EN
\EQ
S=\sum_{m=0}^{\infty} \int_{-\infty}^{\infty}
\left [(\rho_m(x) +\tilde{\rho_m}(x))ln(\rho_m(x) +\tilde{\rho_m}(x))
-\rho_m(x) ln(\rho_m(x)) -\tilde{\rho_m}(x) ln(\tilde{\rho_m}(x)) \right ]
\EN
where we made use of a simplified notation, $\sigma(x)~(\tilde{\sigma}(x))
\equiv \rho_0(x)~(\tilde{\rho_0}(x))$.

Using $\delta F =0$, we find the following equations
\begin{eqnarray}
\chi(\beta) &= & \frac{m}{T} ch(\beta) -\Psi_1^{\prime}(\beta-\beta^{\prime})*
ln(1+e^{-\chi(\beta^{\prime})})
-\sum_{m=1}^{\infty} \Psi_m^{\prime}(\beta-\lambda)*
ln(1+e^{-\epsilon_m(\lambda)})
\end{eqnarray}
\begin{eqnarray}
ln(1+e^{\epsilon_m(\lambda)})= \Psi_m^{\prime}(\lambda-\beta^{\prime})*
ln(1+e^{-\chi(\beta^{\prime})})
+\sum_{k=1}^{\infty} A_{km}(\lambda-\lambda^{\prime})*
ln(1+e^{-\epsilon_k(\lambda^{\prime})})
\end{eqnarray}
where we have redefined the densities
$\frac{\sigma}{\tilde{\sigma}}=e^{-\chi}$;
$\frac{\rho_m}{\tilde{\rho_m}}=e^{-\epsilon_m}$. The symbol $g(x-y)*f(y)$
denote
the convolution $\frac{1}{2 \pi} \int_{-\infty}^{\infty} g(x-y) f(y) dy$;~
$\Psi_m^{\prime}(x)= \frac{d}{dx} \Psi_m$ and $A_{mk}(x)=
\frac{d }{dx} \Theta_{mk}(x) +2 \pi \delta_{m,k} \delta(x)$.

In order to simplify the Eqs.(14,15) it is important to use the following
relations,
\EQ
A_{mk}^{-1}(x)= 2 \pi \delta_{m,k} \delta(x) -\phi(x)
(\delta_{m+1,k}+ \delta_{m-1,k})
\EN
\EQ
\Psi_m^{\prime}(x)= \phi(x-y)*A_{m2}(y);~~ \phi(x)=\frac{1}{ch(x)}
\EN

First applying $A_{km}^{-1}$ in Eq.(15) and eliminating the sum in Eq.(14)
by using Eq.(15), we finally obtain the following result,
\EQ
\chi(\beta) +\phi(\beta-\lambda)*ln(1+e^{-\epsilon_2(\lambda)}) =
\frac{m}{T}ch(\beta)
\EN
\EQ
\epsilon_m(\lambda)
+\phi(\lambda-\lambda^{\prime})* \left [
ln(1+e^{-\epsilon_{m+1}(\lambda^{\prime})}) +
ln(1+e^{-\epsilon_{m-1}(\lambda^{\prime})}) \right ] +
\delta_{m,2} \phi(\lambda-\beta)*ln(1+e^{-\chi(\beta)})=0
\EN
in accordance with the results conjectured in ref. \cite{faza}.

It is also interesting to study the model (1) in presence of a magnetic
field h coupled with one of the components of the conserved charge
$\vec{Q}=\int \vec{\eta} X \partial_t \vec{\eta} d^2 x$. Now, the energy
changes to $E=\sum_{j=1}^{N}m ch(\beta_j) +h(N-M)$ and the minimization
procedure follows close to the one for the Heisenberg chains \cite{str,taba}.
In this case, we have an extra boundary condition
to the variables $\epsilon_m$, such that
$\lim_{m \rightarrow \infty} \frac{\epsilon_m}{m} =h$. Strictly at T=0, the
rapidities $\lambda_j$ are frozen, and one can reproduce the known results
obtained by Wiegmann \cite{wi}.

It will be interesting
to generalize our results to the general O(N) $\sigma$-model.
In doing that, we might generate some new TBA equations for perturbed
conformal field theories, by projecting out the strings at some finite
value m. Such approach has successfully been applied
in the case of the critical RSOS model
\cite{ba}. Based on ref. \cite{faza} and our present
result this approach  can also be used in the
general O(N) $\sigma$-models.



\end{document}